\documentclass[lettersize,journal]{IEEEtran}
\usepackage{amsmath,amsfonts}
\usepackage{algorithmic}
\usepackage{algorithm}
\usepackage{array}
\usepackage{color}
\usepackage{placeins}
\usepackage{float}
\usepackage{tabularx,colortbl}
\usepackage[caption=false,font=normalsize,labelfont=sf,textfont=sf]{subfig}
\usepackage{textcomp}
\usepackage{stfloats}
\usepackage{url}
\usepackage{verbatim}
\usepackage{graphicx}
\usepackage{cite}
\begin{document}

\title{Interplay Between AI and Space-Air-Ground Integrated Network:  The Road Ahead}

	\author{Chenyu Wu, \IEEEmembership{Member, IEEE},  Xi Wang, Yi Hu, Shuai Han, \IEEEmembership{Senior Member,~IEEE},
	\\	Weixiao Meng, \IEEEmembership{Senior Member,~IEEE}, and Dusit Niyato, \IEEEmembership{Fellow,~IEEE} 
	
	\thanks{
	
	This work of C. Wu was supported in part by the National Natural Science Foundation of China under Grant number 62401175, in part by the Postdoctoral Fellowship Program of CPSF under grant number GZC20233483, and in part by the China Postdoctoral Science Foundation under Grant number 2024M764187.

	C. Wu, X. Wang, Y. Hu, S. Han, and W. Meng are with the School of Electronics and Information Engineering, Harbin Institute of Technology, Harbin 150001, China (e-mail: wuchenyu@hit.edu.cn; wangxi\_chn@foxmail.com; huyimiaomiao@163.com; hanshuai@hit.edu.cn; wxmeng@hit.edu.cn).

	Dusit Niyato is with the College of Computing and Data Science, Nanyang Technological University,
	Singapore 639798. (email: dniyato@ntu.edu.sg)
		
	\textit{(Corresponding author: Shuai Han.)}
	%
		
}

}



\maketitle

\begin{abstract}
Space-air-ground integrated network (SAGIN) is envisioned as a key network architecture for achieving ubiquitous coverage in the next-generation communication system. Concurrently, artificial intelligence (AI) plays a pivotal role in managing the complex control of SAGIN, thereby enhancing its automation and flexibility. Despite this, there remains a significant research gap concerning the interaction between AI and SAGIN. In this context, we first present a promising approach for developing a generalized AI model capable of executing multiple tasks simultaneously in SAGIN. Subsequently, we propose a framework that leverages software-defined networking (SDN) and AI technologies to manage the resources and services across the entire SAGIN. Particularly, we demonstrate the real-world applicability of our proposed framework through a comprehensive case study. These works pave the way for the deep integration of SAGIN and AI in future wireless networks.

\end{abstract}


\section{Introduction}
\IEEEPARstart{W}{ith} the emergence of vertical industries such as low-altitude economy, commercial aerospace, and autonomous driving, the sixth generation (6G) communication system is anticipated to support ubiquitous connectivity, ultra-high speeds, and enhanced intelligence to accommodate new applications. To fulfill these requirements, it is essential to implement new network architectures and flexible paradigms that can overcome the limitations of terrestrial wireless networks. Space-air-ground integrated network (SAGIN), which incorporates geostationary-earth orbit (GEO) satellites, low-earth orbit (LEO) satellites, and aerial platforms (such as unmanned aerial vehicles (UAVs)) with terrestrial segments, is a key enabling network architecture for meeting the seamless connectivity demands of 6G\cite{qian}. SAGIN leverages the advantages of different network layers to achieve synergy, and thus can respond to diverse user activities in modern communication networks, especially in remote areas or densely populated areas.

In recent years, wireless networks have witnessed an upsurging trend towards self-organization and automation, with the persistent evolution of artificial intelligence (AI) technologies. It is undeniable that AI will be integrated and standardized across all open systems interconnection (OSI) layers of 6G networks, fostering a symbiotic interaction among them\cite{6G}. Due to the inherent properties of SAGIN, such as heterogeneous structure, distributed resources, and dynamic topology, the configuration and management of SAGIN are generally more complicated than terrestrial networks. In this context, researchers have begun to explore the use of AI in various applications within SAGIN including spectrum sharing, radio resource scheduling, traffic offloading\cite{offload}, and routing\cite{kato}, etc.  On the other hand, significant progress has been made in the application of large AI models across various domains in these years. For example, in natural language processing (NLP), foundation models like GPT-4, PaLM, and DeepSeek have demonstrated remarkable capabilities in text generation, translation, and reasoning. However, when it comes to the specialized field of optimizing SAGIN, there remains a substantial research gap.

Although incorporating AI in SAGIN can yield substantial performance enhancements, it still encounters the following challenges related to architectural flexibility and availability:

\begin{itemize}
	\item \textbf{From the optimization perspective}: AI technologies such as deep
	learning (DL) and deep reinforcement learning (DRL) have shown significant potential in addressing decision-making and optimization problems\cite{10599519}. However, existing AI frameworks predominantly focus on specific tasks, resulting in a lack of universality and sustainability. Besides, training a dedicated AI model for each specific task necessitates huge training costs.   	
	\item \textbf{From the network management perspective}: Traditional networks typically employ fixed architectures and  management approach, which lack agility and flexibility\cite{automation}. Software-defined networking (SDN) and network function virtualization (NFV) represent promising technologies for achieving flexible network management and reducing operation expenses. These technologies originated from wired service providers and have been recently expanded to wireless networks. However, 
	implementing SDN and NFV in SAGIN for autonomous management still faces critical challenges due to the diverse service requirements, dynamic topology, and cross-domain differences. 

\end{itemize}

To address these challenges, this paper provides a comprehensive overview of the role of AI in next-generation SAGINs. The main contributions are summarized as follows.
\begin{itemize}
	\item  We review the application of traditional AI methods in typical task execution scenarios within SAGIN and point out their limitations. We introduce the incorporation of large generative AI models into our proposed framework, designed to address a diverse range of SAGIN applications in a unified manner.
	\item Leveraging the elasticity of SDN and NFV, we design a multi-domain framework for managing SAGIN. Within this framework, we introduce  how AI techniques can effectively schedule the cross-domain service function chain under dynamic topology. Through case studies, the effectiveness of this framework in practical scenarios has been verified.
\end{itemize}


\begin{table*}[t]
	\centering
	\caption{List of Acronyms}
	\label{tab:acronyms}
	\footnotesize
	\setlength{\tabcolsep}{3pt}
	\begin{tabular}{|l|l|l|l|l|l|} 
		\hline
		\textbf{Acronym} & \textbf{Full Name} & 
		\textbf{Acronym} & \textbf{Full Name} & 
		\textbf{Acronym} & \textbf{Full Name} \\
		\hline
		A3C & Asynchronous Advantage Actor-Critic & 
		DRL & Deep Reinforcement Learning & 
		LEO & Low Earth Orbit \\
		AI & Artificial Intelligence & 
		DNN & Deep Neural Network & 
		LSTM & Long Short-Term Memory \\
		BAIM & Big AI Model & 
		eMBB & Enhanced Mobile Broadband & 
		mMTC & Massive Machine-Type Commun. \\
		CSI & Channel State Information & 
		GAN & Generative Adversarial Network & 
		NC & Network Calculus \\
		DL & Deep Learning & 
		GAT & Graph Attention Network & 
		NFV & Network Function Virtualization \\
		\hline
		GEO & Geostationary Earth Orbit & 
		NLP & Natural Language Processing & 
		SAGIN & Space-Air-Ground Integrated Network \\
		GenAI & Generative AI & 
		OSI & Open Systems Interconnection & 
		SDN & Software-Defined Networking \\
		HAP & High-Altitude Platform & 
		QoS & Quality of Service & 
		SFC & Service Function Chain \\
		RF & Radio Frequency & 
		SLA & Service Level Agreement & 
		TAG & Temporal Aggregation Graph \\
		UAV & Unmanned Aerial Vehicle & 
		URLLC & Ultra-Reliable Low-Latency Commun. & 
		VNF & Virtual Network Function \\
		\hline
	\end{tabular}
\end{table*}
\section{AI Applications for SAGIN Optimization}

In this section, we first overview existing AI algorithms in addressing the limitations of conventional optimization approaches in various application scenarios of SAGIN. We then highlight the limitations of these distributed AI algorithms and introduce the research direction of devising a generalized AI framework based on big generative AI model to realize the integration of multiple functions. Besides, several challenges and opportunities are discussed to inspire future research.

\subsection{Representative Use Cases}\label{sec_opt}

\begin{figure*}[!t]
	\centering
	\includegraphics[width=0.95\textwidth]{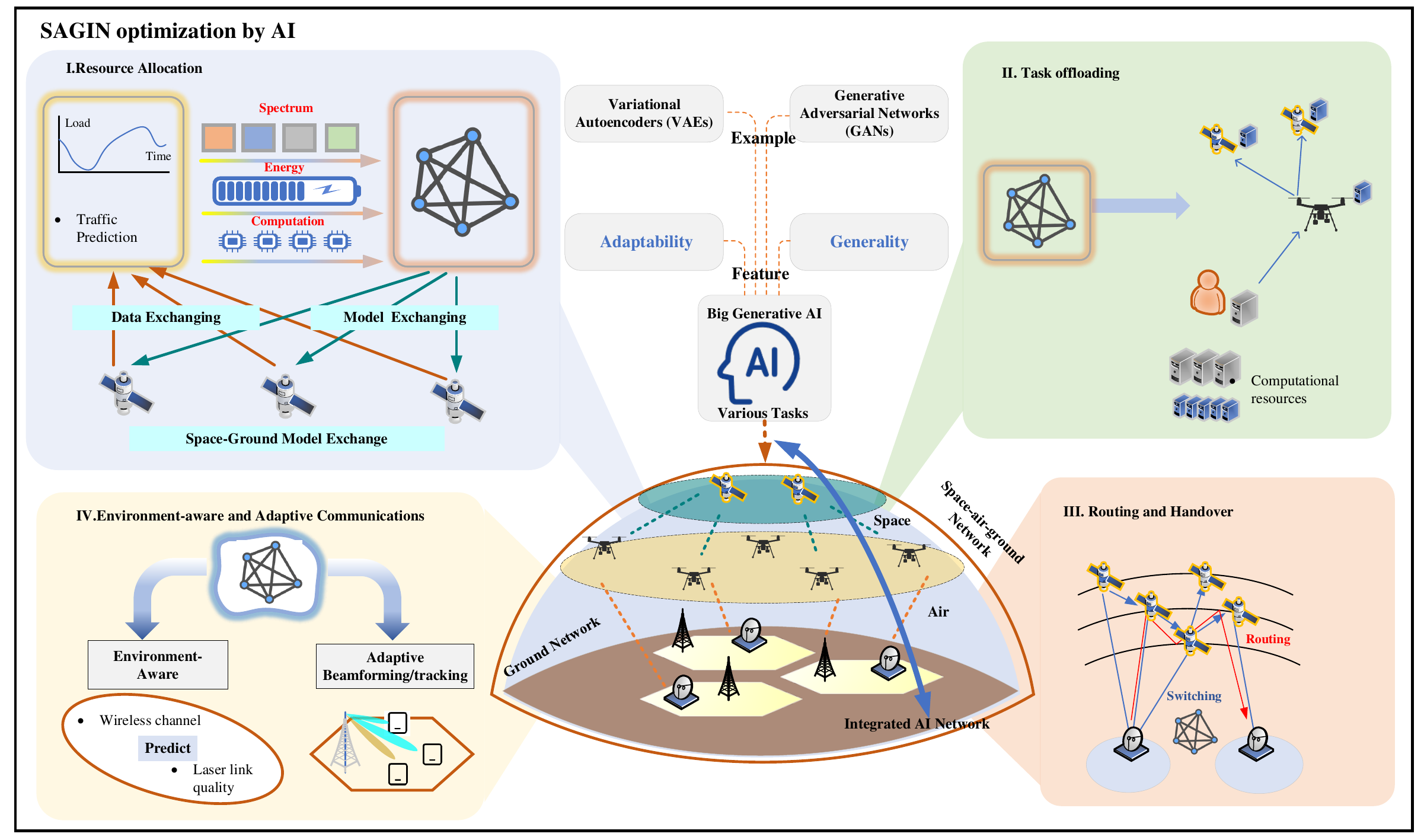}
	\caption{A generalized framework based on big generative AI for tackling various tasks in SAGIN. Representative use cases include resource allocation, task offloading, routing, and environment-aware and adaptive communication.}
	\label{fig_optimization}
\end{figure*}

Due to the multi-layer heterogeneous architecture and complex environmental parameters, AI technology plays an increasingly important role in optimizing SAGIN\cite{offload,kato,10599519}. 
Through advanced AI technologies such as DRL, federated learning, and edge intelligence, SAGIN can achieve dynamic optimization, real-time response, and adaptive adjustment in complex and multi-layered networks\cite{10103768}. In new scenarios such as intelligent disaster response, smart city management, and low-altitude economy, AI technologies provide SAGIN with more intelligent and efficient solutions, driving the global communication network to new heights. We introduce in the following the representative application scenarios in SAGIN where AI is indispensable, as illustrated in Fig. \ref{fig_optimization}.

\subsubsection{Resource Allocation} 
SAGIN possesses multi-dimensional resources including communication-related resources, computation-related resources, and energy. Compared with terrestrial systems, the spectrum in space is more scarce, potentially leading to severe interference with base stations. DL architectures such as long short-term memory (LSTM) can be utilized for analyzing
information of data and the geographic location of users, which can be leveraged for 
efficient spectrum allocation across different layers and nodes. LEO satellites’ high-speed motion introduces rapid beam handovers and inter-satellite interference. For multi-satellite beam allocation, graph neural networks can model satellite topology dynamics, while multi-agent DRL optimizes beam coordination to minimize overlap and maximize coverage. For instance, LSTM-based predictors can anticipate user-satellite visibility windows, enabling proactive beam switching across constellations.

\subsubsection{Task Offloading} 
Satellites and air nodes are potential candidates to complement the task offloading of terrestrial nodes, constituting a hierarchical offloading structure. Nevertheless, due to the dynamic nature of topology as well as service requirements, conventional optimization strategies are no longer suitable for SAGIN. To address this issue, DL can be utilized to analyze historical data, predict network congestion and computational load, and assist in advance with task scheduling decisions. Besides, DRL algorithms can identify an optimal task offloading solution among multiple options by dynamically exploring different combinations of strategies. This process can be completed offline, thus facilitating a rapid response to online decisions. In this context, the authors in \cite{offload} 
proposed a double Q-learning algorithm with delay-sensitive replay memory to determine the offloading strategy based on historical information. However, this model only depicts a snapshot of the offloading scenario and cannot adapt to dynamic channels and varying tasks.

\subsubsection{Routing} Routing refers to the process of selecting the optimal path for data delivery through various layers.  AI plays a pivotal role in assisting in solving routing and link switching problems within SAGIN. By analyzing real-time data on network conditions, link qualities, and traffic patterns, AI can predict and adapt to network dynamics, enabling more intelligent and efficient routing decisions. For instance, AI can identify potential bottlenecks or disruptions in advance (e.g., when satellites are moving across the polar region) and reroute data packets through alternative paths. 

\subsubsection{Environment-Aware and Adaptive Communications}

The incorporation of AI into SAGIN offers a pioneering approach to environmental perception and adaptive communication. For instance, the authors of \cite{xingwang} addressed the joint beamforming issue using DRL in a SAGIN context, where satellites communicate with high-altitude platforms (HAPs) via laser links, and HAPs interact with the ground through radio frequency (RF) links. In reality, while laser links are efficient, they are more susceptible to atmospheric attenuation and obstruction, compared with RF links. When SAGIN's links support both laser and RF transmission, AI can proactively predict the quality of laser links in real time. This capability allows the system to dynamically switch or modify communication paths to RF links as laser link conditions deteriorate, thereby ensuring consistent and reliable connectivity throughout the network. Moreover, in satellite communications, AI algorithms can be utilized for adaptive beamforming and beam tracking to dynamically adjust beam direction and shape, thereby enhancing link quality and user experience.

\subsection{Design Guideline and Challenges}
The application of AI algorithms in SAGIN can significantly enhance network performance and efficiency. However, a prevalent challenge is that most existing AI solutions are task-specific, lacking the generality and adaptability to address the diverse and dynamic challenges inherent in SAGIN. To overcome these limitations, the adoption of big AI models (BAIM) \cite{zhaohui} and generative AI (GenAI) \cite{dusit} offers promising solutions. BAIMs are designed to capture complex patterns and relationships across diverse data sources, thereby improving their applicability to a wide range of tasks. GenAI, on the other hand, can generate content by leveraging the statistical properties of the input data. Therefore, big generative AI model can be used for executing the tasks mentioned above to enhance the extensibility and versatility of SAGIN operations.

For instance, the accurate feedback of channel state information (CSI) and precise localization in SAGIN rely heavily on the intricate feature extraction of pilot signals. By integrating these functionalities within a single BAIM, the model can learn to correlate the features of pilot signals with both CSI and positioning metrics, enabling compressed signal size and intelligent inference\cite{zhaohui}. Moreover, advanced GenAI techniques like generative adversarial network (GAN) can aid in CSI estimation and resource allocation in SAGIN through
producing virtual scenarios by generator and refining these generations by discriminator\cite{dusit}.  Another noteworthy application is within the scope of predictive maintenance and fault management.  BAIMs are capable of analyzing historical and real-time data from various network components to forecast potential failures or performance degradation. This predictive capability  enables preemptive maintenance actions, thereby decreasing downtime and enhancing the overall reliability of the network. For example, BAIMs can anticipate the degradation of satellite components or the probability of drone malfunctions based on environmental conditions and usage patterns.

A big generative AI framework can be deployed to integrate the functions mentioned Section \ref{sec_opt}. As illustrated in Fig. \ref{fig_optimization}, the training of such a big generative AI model typically involves a cloud-edge collaboration framework. Specifically, the model is initially pre-trained in cloud servers, e.g., GEO satellites, while the pre-trained model is fine-tuned at the edge, tailored to specific tasks or local network conditions. Firstly, the amount of data required to train and validate the model is enormous. The distributed nature of SAGIN makes data collection and management more difficult. Secondly, the asymmetric network characteristics (e.g., transmission latency and computational resources) across different segments make it hard to implement a huge AI model in real-time. Finally, guaranteeing the consistency and convergence of the model in such a heterogeneous and dynamic environment is non-trivial.

In order to surmount these challenges, it is imperative for researchers to persistently devise novel algorithms and optimization techniques. These should be capable of managing large-scale data processing, accommodating asymmetric network conditions, and ensuring model performance in real-world deployments.

\section{AI Applications for SAGIN Management}

The management of SAGIN (including network resources and traffic) is typically entrusted to service providers. The advent of SAGIN presents significant challenges to traditional network management methods. 
This section focuses on the orchestration of service function chains (SFCs) in SAGIN, and proposes an intelligent management framework named AI-based SFC orchestration (AI-SFCO). We first introduce the framework of AI-SFCO and then answer \textit{1) how AI techniques can guarantee the end-to-end service performance of SFCs; 2) how to adapt to the dynamic topology changes induced by SAGIN features; 3) how to coordinate the orchestration of SFCs across multiple domains}. In addition, we demonstrate through a specific simulation case that AI technology in our framework can significantly improve the success rate and revenue of deploying SFCs in a dynamic SAGIN environment.

\subsection{Framework of AI-SFCO}

SFCs refer to a series of network service functions, such as firewalls, intrusion detection systems, and load balancers\cite{bib1}. These functions are executed in a specific order within the network to ensure that traffic undergoes necessary optimization measures. SFCs are utilized in a broad range of applications, from traffic management inside data centers to guarantee quality of service (QoS) in wide area networks. The study of SFC deployment is crucial for effective network service management, which encompasses several key aspects. Firstly, the increasing growth of network traffic and the trend toward service diversification make traditional static service deployment methods inadequate for modern networks. Secondly, the deployment of SFCs must take into account multiple factors such as latency, bandwidth, and security, which often conflict with each other.\cite{bib2}. Thirdly, the introduction of the SAGIN brings a more dynamic and complex network environment, in which the mobility and connectivity uncertainty of nodes increase the difficulty of SFC deployment\cite{bib3}. 


\begin{figure*}[!t]
	\centering
	\includegraphics[width=7in]{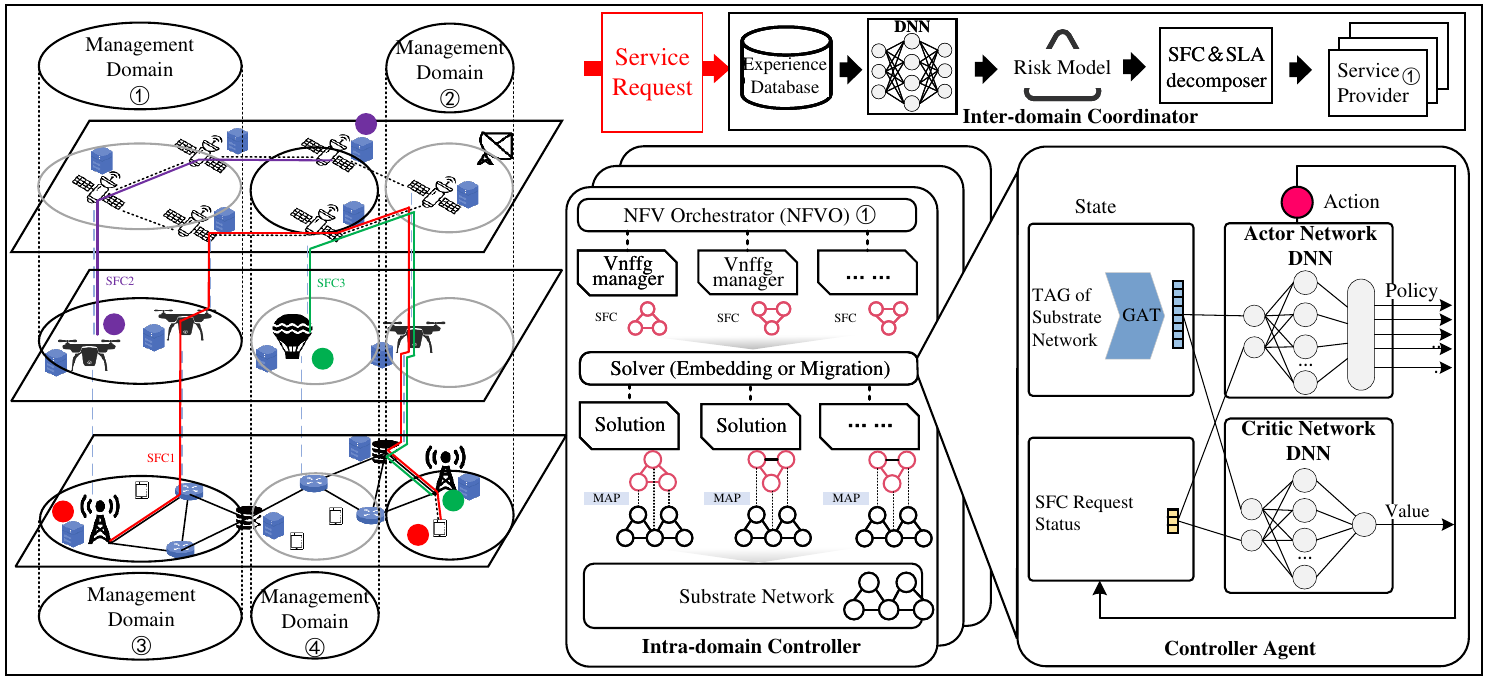}
	\caption{AI-SFCO: the framework for managing multi-domain SAGIN. The intra-domain controller handles the SFC orchestration according to the network status, while the inter-domain controller is responsible for the communication and collaboration between different management domains.}
	\label{fig_Framework}
\end{figure*}
\begin{table*}[!t]
	\caption{Classification and Summary of literature related to network orchestration technology\label{tab_Survey}}
	\centering
	\begin{tabular}{|l||l||p{3cm}||p{9cm}|}
		\hline
		Category & Reference & Optimization & Contribution \\
		\hline
		U/F/S & \cite{bib1} T. Wang et al. (2023) & DRL & Joint admission control and resource allocation policies\\
		C/F/S & \cite{bib2} J. Pei et al. (2020) & DL & Improve network QoS and user experience\\
		U/M/S & \cite{bib3} V. Eramo et al. (2017) & Reinforcement Learning & The energy consumption of the server to open the VNF and the information loss during the SFC migration are considered\\
		C/M/S & \cite{wangxi} X. Wang et al. (2024) & Meta-heuristic & Resource-aware SFC embedding and migration scheme with low time complexity.\\
		C/F/D & \cite{bib6} C. Wang et al. (2022) & DRL & SAGIN memory resource management algorithm for distributed DRL\\
		\hline
	\end{tabular}
\end{table*}

In view of this, we design an intelligent management architecture based on the combination of AI and SDN/VNF to deal with the above challenges, as shown in Fig. \ref{fig_Framework}. We divide the entire SAGIN into several \textit{management domains}, which can be defined according to different operators or geographical regions. Each management domain contains its own network components including ground nodes, UAVs, and GEO and LEO satellites, and exhibits unique network characteristics. For example, one management domain may prioritize high-density connectivity, while another may need to cope with the high latency issues arising from long distance communications.

Inside each administrative domain, intelligent \textit{intra-domain controllers} based on AI are deployed to monitor the network status in real time and dynamically adjust the SFCs according to the current network status. For example, within a data center management domain, intelligent controllers can be deployed on core switches or data center servers (e.g., GEO satellites) to achieve centralized management and rapid response. Within a radio access management domain, intelligent controllers may be located at base stations or edge computing nodes (UAVs and LEOs) to ensure real-time decision-making capabilities. Moreover, to orchestrate the SFCs across domains, we deploy an AI-driven \textit{inter-domain coordinator} in the network, which is responsible for the collaboration between different management domains to ensure that the SFC operates seamlessly across multiple domains. It is usually located on cloud platforms or high-performance computing centers to enable efficient data processing and decision making.

The controller in each domain periodically reports the state information of the local network to the inter-domain coordinator, including but not limited to traffic patterns, network load, and QoS metrics. Upon receiving these information, the inter-domain coordinator employs advanced AI algorithms to evaluate the risk of providing reliable services after SFC deployment. The aim is to optimally partition the SFC to fulfill the cross-domain deployment request, and send the optimized strategy to the controllers in each domain. The intra-domain controller adjusts the local service function chain configuration according to the newly received policy, thereby achieving the global optimal service deployment. This closed-loop feedback mechanism ensures that the network can adapt to changing conditions while maintaining optimal QoS.

\subsection{Challenges and Solutions}

When deploying SFCs in SAGIN, the main challenges include guaranteeing end-to-end service performance, real-time migration caused by dynamic network topology changes, and orchestration of SFCs across multiple management domains. We review research studies in the direction of SFC orchestration and select representative studies in Table \ref{tab_Survey} according to whether to consider end-to-end performance constraints (\textbf{U}nconstrained\textbar \textbf{C}onstrained), whether to use dynamic migration strategy (\textbf{F}ixed \textbar\textbf{M}igratable), and whether to consider cross-domain orchestration (\textbf{S}ingle-Domain\textbar  Multi-\textbf{D}omain). In the following, we first analyze the above three challenges faced by providing network services in SAGIN, and then give our design in the AI-SFCO framework for tackling them.

\subsubsection{Service Performance Guarantee}

Deploying SFC in SAGIN presents significant challenges arising from the complexity of resource allocation, network environment heterogeneity, and service function diversity. Firstly, different types of network nodes possess distinct computing, (transmission) bandwidth and storage resources. The allocation of these resources necessitates consideration of fluctuations during actual operation to ensure a predictable end-to-end service guarantee. Secondly, SAGIN encompasses a range of communication protocols, each with varying transmission rates, coverage, and reliability, thereby escalating the complexity of service performance assurance. Finally, different service functions have unique resource requirements, which necessitate sophisticated refined control strategies to maintain the overall service performance.

In order to ensure consistent service performance in SAGIN, the authors of \cite{bib2} employed a deep belief network to extract and learn the features of node and link attributes in the network. Two networks, namely the VNF selection network and the VNF chaining network, were used to embed SFCs in steps to guarantee the QoS of users. However, the fixed resource allocation methods in \cite{bib2} can not adapt to user dynamic business requirements. To address this shortcoming, the intra-domain controller in our AI-SFCO framework combines the resource characteristics of nodes and links with DRL and network calculus (NC) theory to achieve fine control over network resources. Particularly, NC aids in modeling and analyzing the delay performance of SFCs. Moreover, through a novel framework of DRL, i.e., asynchronous advantage actor-critic (A3C), the intra-domain controller is endowed with self-learning and self-optimization capabilities. Specifically, the agent (mapping of the intra-domain controller) can monitor the network status in real-time and dynamically adjust the priority of the SFC in accordance with the real-time network status to meet the business performance requirements of critical tasks. 


\subsubsection{Dynamic Topology Adaption}

A significant challenge in deploying SFCs in SAGIN is managing the real-time migration caused by the dynamic changes of the network topology. Specifically, the connection relationships between network nodes such as space-borne equipment and air-borne equipment move or change with the trajectory of the carrier platform, resulting in frequent updates of the network topology. These dynamic changes require the SFC to be capable of migrating in real time to maintain service continuity and efficiency. Hence, it is necessary to optimally allocate the network resources while maintaining the stability of the link during the migration process to ensure that the performance of the SFC is not affected.

To adapt to dynamic topology, our prior work \cite{wangxi} proposed a resource-aware meta-heuristic algorithm to handle the burstiness of UAV network topology changes during SFC deployment. However, there remains a large room for algorithm improvement due to the regularity of topology change in SAGIN. Specifically, the topology changes can be predicted based on data such as satellite orbit parameters and UAV trajectory planning. In AI-SFCO, the intra-domain agent can combine the temporal aggregation graph (TAG) model with the graph attention network (GAT) to predict dynamic topology of SAGIN, and migrate the service functions in real time based on the prediction results.  Learning from dynamic graph features, the agent decides when and how to migrate service functions. For example, when the agent predicts that a satellite is about to enter the Earth's shadow area, which may cause communication disruption, it can direct the migration of relevant service functions to an alternative satellite.

\subsubsection{Cross-Domain Coordination}

It is unrealistic to manage the huge system of SAGIN only relying on a single control plane. In practice, the coordination of multiple management domains is essential for the orchestration of SFCs since different management domains have their own specific constraints and service level agreement (SLA) requirements. For instance, one domain may prioritize high-bandwidth and low-latency services, while another might emphasize security and redundancy. The orchestration of cross-domain SFCs must ensure end-to-end QoS and sustain a consistent service level, even with fluctuating network conditions and business needs. These challenges require that the SFC possesses a high degree of flexibility and intelligent decision-making capability when deploying SFCs across domains.

In order to manage SFCs in SAGIN across multiple domains, the authors in \cite{bib6}  proposed a scheme based on distributed DRL. In our AI-SFCO framework, we delegate the authority of SFC deployment to the service provider rather than the user. Particularly, an inter-domain coordinator is integrated to guarantee cross-domain service performance within SAGIN. The process is detailed as follows. First, the inter-domain coordinator evaluates the network capabilities of each management domain by consulting historical records stored in the database. Then, it employs the deep neural networks (DNN) to assess the resource status and service demands of each domain, and determines the risk model associated with deploying SFCs with varying requirements across different domains. Finally, it dynamically modifies the decomposition strategy of both SFC and SLA to ascertain the optimal allocation of SFC resources among domains. Through continuous accumulation of data in the experience database and DNN training, the coordinator refines its ability to anticipate network condition shifts, thereby preemptively adjusting the SFC decomposition strategy. For instance, the coordinator can forecast resource demands during peak hours as well as whether the resources of a certain domain are about to be saturated, and then pre-allocate adequate resources to mitigate potential service degradation. 

\subsection{A Case Study}

\begin{figure}[!t]
	\centering
	\includegraphics[width=3.5in]{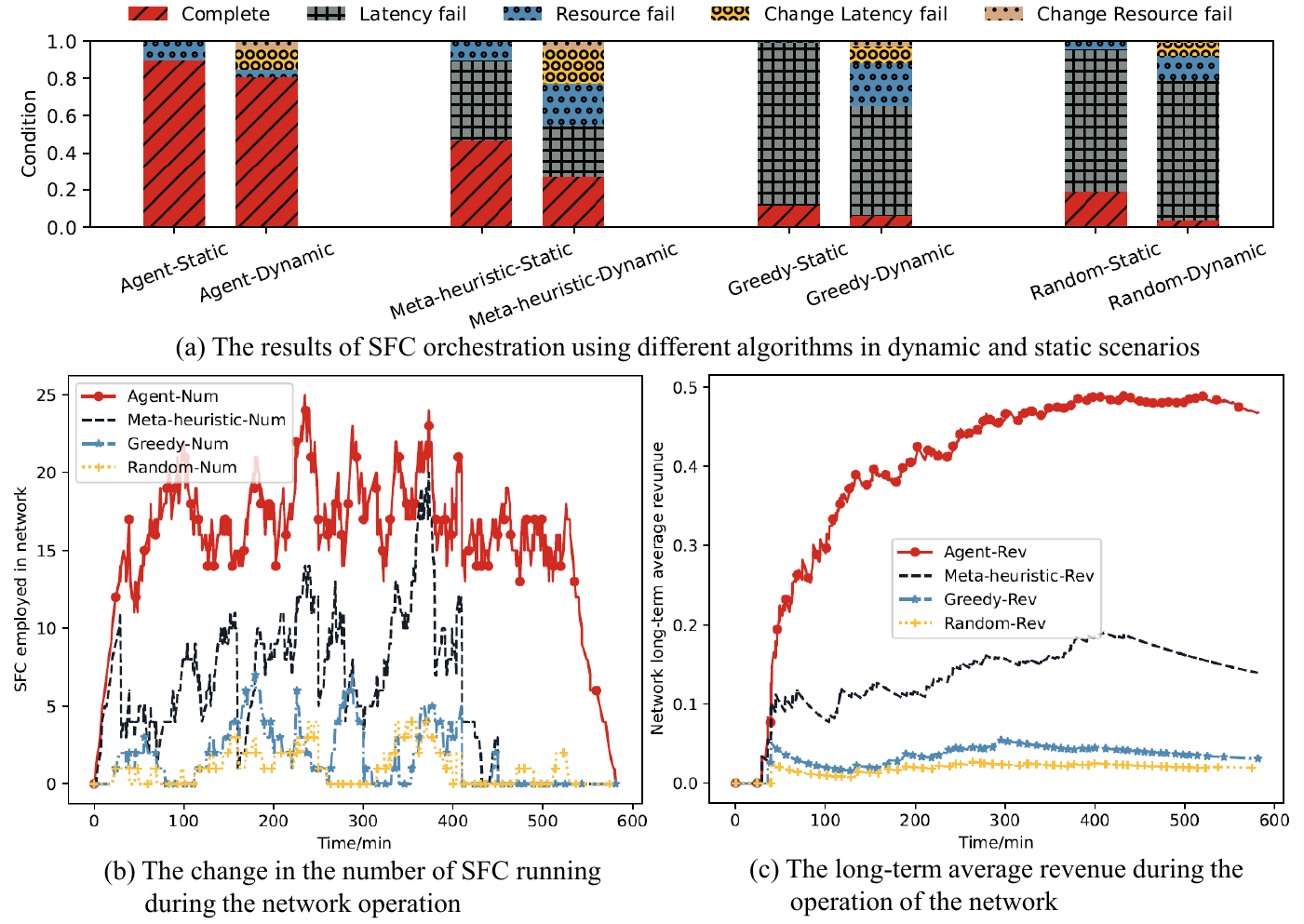}
	\caption{Simulation results for the deployment of SFCs in the SAGIN. Key A3C-related parameters are summarized as follows: Number of actors: 4; learning rates of the actor and critic networks: 0.0025 and 0.0005, respectively; discount factor for temporal difference error: 0.95; batch size for experience replay: 64; number of hidden units per layer: 64.}
	\label{fig_CaseStudy}
\end{figure}


In this section, we present a concrete simulation case \footnote{Our open-source platform: \url{https://wangxichn.github.io/mini_sfc/}} to demonstrate how our designed intra-domain controller is utilized to deploy SFCs in SAGIN for adapting to the changing topology and realizing dynamic migration. We use DRL and GAT to extract the features of link reliability, and then complete the selection of SFC deployment locations. Specifically, DRL can be used to learn the policy to make the best decision when the network state changes, while GAT is able to capture the dependencies between nodes in the network and extract information useful for decision making from them. In this way, the system can automatically learn how to dynamically adjust the deployment of the SFC according to the current state and demand of the network.

\textit{We simulate a real event to validate our design is suitable for real-world deployment}. The simulation scenario is inspired by the catastrophic floods that hit Henan Province of China in 2021, where communication infrastructures were severely destroyed. To alleviate the impact of the disaster, we assume a SAGIN is deployed for disaster relief. Four orbits of the Kuiper constellation are set at an altitude of 590 km. 10 satellites are uniformly distributed in each orbit to form the satellite network layer. Additionally, five UAVs are placed on preplanned trajectories to serve as temporary communication nodes over the affected areas, while three surviving ground base stations offer basic connectivity. These elements mimic the real-world scenario during a disaster: satellites offer wide-area coverage, UAVs function as mobile relays, and ground stations anchor local communication.

In terms of computational resources, space, air, and ground nodes are equipped with capacities of 3 Gb/s, 300 Mb/s, and 20 Gb/s, respectively. All nodes are furnished with 512 Gb of memory. The bandwidths for space-ground and inter-satellite links are set as 200 Mbps and 500 Mbps, respectively. To capture dynamic changes in connectivity, the simulation operates for a duration of 10 hours (from 4:00 to 14:00), taking 60 snapshots of the network topology at 10-minute intervals.  This configuration ensures both robustness and adaptability, enabling SFCs tailored for URLLC, mMTC, and eMBB applications.

To train the model, we employ the A3C ``master-worker" parallel architecture. This architecture allows multiple agents (workers) to simultaneously interact with the environment, collect empirical data, and send this data to a central agent (master) to learn and update policies. In this way, the agent is able to learn faster the policy that makes the best decision when the network state changes. In order to verify the effectiveness of the proposed intelligent method, the following three methods are selected for comparison: 1) \textit{Meta-heuristic}\cite{wangxi}: based on the meta-heuristic algorithm of global resource capabilities, VNFs are mapped to physical nodes; 2) \textit{Greedy}: preferentially select nodes with sufficient resources; 3) \textit{Random}: random mapping of nodes that meet the resource requirements.

The experimental results are shown in Fig. \ref{fig_CaseStudy}. We compare the network service orchestration of dynamic environment deployment and static environment deployment using one of the topology snapshots selected. We also counted various reasons for service completion and non-completion throughout the entire simulation process, as illustrated in Fig. \ref{fig_CaseStudy}(a). Fig. \ref{fig_CaseStudy}(b) depicts the number of active service function chains (SFCs) in the network during operation, while Fig. \ref{fig_CaseStudy}(c) shows the long-term average revenue of the network during service delivery, calculated as the ratio of completed services to time.

By synthesizing these results, it is evident that the proposed AI-based framework significantly outperforms traditional embedding strategies. This improvement stems from the ability of the proposed framework to leverage the dynamic characteristics of link changes in SAGINs. Specifically, the AI-driven approach dynamically adapts to varying network conditions, such as frequent topology changes caused by satellite motion, UAV repositioning, and fluctuations in ground station availability. Through intelligent prediction and response to these changes, the framework ensures more efficient SFC placement and routing, thereby enhancing overall performance.

Furthermore, the results imply that the scalability of this framework makes it suitable for large-scale and diverse SAGIN scenarios. In such scenarios with complex and uncertain network dynamics, effectively leveraging temporal and spatial patterns is the key to improving performance.  The proposed framework can not only increase service completion rates but also enhance resource utilization, thereby achieving higher long-term average revenue (as shown in Fig. \ref{fig_CaseStudy}(c)). The proposed framework has shown its practical utility in ensuring dependable communication in the context of large-scale disasters. This indicates that our framework could potentially be expanded to accommodate increasingly intricate and heterogeneous SAGIN configurations, positioning it as a promising solution.

In summary, the performance improvements for SFC orchestration is attributed to the framework’s ability to intelligently exploit network dynamics.  This is particularly important in large-scale or diverse SAGIN scenarios, where traditional approaches may not be able to provide an efficient and reliable solution.

\section{Conclusion}
The integration of SAGIN with AI represents a pivotal step towards achieving ubiquitous coverage and enhanced automation in the next generation communication systems. In this paper, we first introduced the applications of AI in executing various tasks within SAGIN and discussed the advantages of developing a generalized AI model capable of executing multiple tasks, including their enhanced scalability and improved efficiency. We then presented a framework, which harnesses SDN and AI technologies to offers a promising solution for managing the complex resources and services across the entire SAGIN. Through a detailed case study, we confirmed the feasibility of our framework in real-world scenarios. 

\bibliographystyle{IEEEtran}
\bibliography{mag_bibliography}

\end{document}